\documentclass{PoS}
\usepackage{epsfig}

\title{Numerical tools for the theoretical study of QCD at small x}

\ShortTitle{Numerical tools for the theoretical study of QCD at small x}

\author{A. Sabio Vera\\
        Physics Department, Theory Division, CERN, CH-1211 Geneva 23, 
        Switzerland\\
        E-mail: \email{Agustin.Sabio.Vera@cern.ch}}

\author{\speaker{P. Stephens}\thanks{This work is 
partly supported by the EU grant mTkd-CT-2004-510126 in partnership with the 
CERN Physics Department and by the Polish Ministry of Scientific Research and 
Information Technology grant No 620/E-77/6.PRUE/DIE 188/2005-2008.}\\
        Institute of Nuclear Physics, Polish Academy of Sciences,\\
        ul. Radzikowskiego 152, 31-342 Cracow, Poland\\
        E-mail: \email{pstephens@annapurna.ifj.edu.pl}}

\abstract{In this contribution we present the status of two numerical 
tools designed to study the small $x$ limit of QCD. The first one is a 
Monte Carlo simulation of the BFKL evolution equation. In design of this 
approach emphasis has 
been placed on exploiting the linear behaviour that many variants of the BFKL 
evolution possess. This allows us to design a procedure which can be used to 
study theoretical and phenomenological aspects of different kernels. The 
second one is a semi-analytic approach to study Lipatov's effective 
action which 
describes Reggeon interactions. The study of the properties of this action is 
very complicated and we propose using a computational tool to handle the large 
amount of non--local vertices and the derivation of higher order corrections.}

\FullConference{DIFFRACTION 2006 - International Workshop on Diffraction in High-Energy Physics \\
		 September 5-10 2006\\
		 Adamantas, Milos island, Greece}

\begin{document}

\section{Introduction}

The structure of Quantum Chromodynamics is very rich. Besides the intrinsic 
uncertainties for the description of the theory in the nonperturbative 
region, the regime where the coupling is small also presents many challenges. 
In order to describe some processes dominated by perturbative dynamics it has 
been important to focus the efforts on different regions of phase space where 
logarithmic factors get enhanced and their resummation to all orders can 
provide 
valuable information. One example of this kind is the Regge limit of 
strong interacting processes. This is a regime where the center of mass 
energy, $s$, is very large and terms of the form $\alpha_s \ln{s}$ are 
enhanced. In processes characterized by two large and similar transverse scales, 
for a range of large energies,  the underlying 
dynamics is linear. This implies that the growth with energy of the 
observables can be described by a linear integral equation, the so--called 
Balitsky--Fadin--Kuraev--Lipatov (BFKL) equation~\cite{FKL}. Linearity has, 
as a 
consequence, an exponential rise of the cross section with an exponent related 
to the so--called hard pomeron intercept. The LO intercept is too large when 
compared to experimental data. The inclusion of the next--to--leading (NLO) 
order 
corrections~\cite{FLCC}, where there is an extra power in the coupling 
when compared to 
the logs of energy, reduces the intercept to values more acceptable from the 
point of view of phenomenology. The BFKL scattering amplitudes contain a 
large amount of physical information which can be fully extracted only by 
numerical methods. It has a Poisson like structure in the number of emissions 
which could be useful in the study of events with a large multiplicity in the 
final state. At NLO one is sensitive to effects like the running of the 
coupling or how to best couple the Reggeized gluons to the external particles. 
These issues can be best treated in a numerical way using the methods 
developed in Ref.~\cite{MC-GGF}. Here we will review some of the equations 
shown in these works and we will introduce a more efficient implementation, 
for alternative ideas see Ref.~\cite{JRA}. The use of the tools shown here
is not limited to the standard forward BFKL kernel but it can be 
extended to the nonforward case which corresponds to colour singlet exchange, 
and more exotic situations as is the case when the QCD evolution is placed 
in a thermal bath.

Although there should be a window of large energies where BFKL effects are 
dominant, linearity leads to the breakdown of unitarity since the Froissart 
bound is not respected for asymptotic energies. This is connected to the 
well--known sentence by the late economist Kenneth E. Boulding: "Anyone who 
believes exponential growth can go on forever in a finite world is either a 
madman or an economist". Therefore at some point unitarity corrections should 
be taken into account. This is a very complicated issue from the theoretical 
point of view and all proposed solutions are always approximations. We 
believe an attractive approach is that proposed by L.~Lipatov in 
Ref.~\cite{Lipatov:1995pn}. A non--linear gauge invariant effective action 
was constructed which describes the interactions between Reggeized quarks 
and gluons with the usual ones. The Reggeized degrees of freedom are 
represented by Wilson lines while the $s$--channel emissions belong to 
a kinematics more general than the quasi--multi--Regge region needed to 
obtain the NLO kernel in the sense that not only a single pair of partons 
is allowed to have a fixed invariant mass but more are allowed. These 
``grouped'' emissions are well separated from each other in rapidity 
following the usual multi--Regge kinematics. This action incorporates 
unitarity in different channels and describes the interaction between 
Reggeons as well as, through quantum fluctuations around the classical 
solution, higher order corrections to the BFKL kernel. In this paper we 
propose the use of a second tool which uses the semi-analytic software 
{\it Effective}~\cite{Hetherington:2006pa} to study and probe the tree--level structure of 
this effective action. In combination with a topology generator, the effective 
vertices of Reggeons coupled to particles can be generated automatically. 

\section{Monte Carlo implementation of BFKL equations}

The first work taking full advantage of the iterative structure present 
in the BFKL equation to develop a Monte Carlo solution for the gluon 
Green's function was presented in Ref.~\cite{Schmidt:1996fg}. More recently 
this method was extended to the much more complicated case of the NLO 
kernel in Ref.~\cite{MC-GGF}. Here we would like to report on recent progress 
on a new implementation of these ideas to continue investigating the 
gluon Green's function in different physical processes.

The LO evolution equation can be written in a very simple form as
\begin{eqnarray}
\frac{\partial}{\partial {\rm Y}} 
{\tilde f} \left(\vec{k}_a,\vec{k}_b,{\rm Y}\right) \equiv \int d^2 \vec{k} \frac{{\bar \alpha}_s }{\pi \vec{k}^2} \theta\left(\vec{k}^2-\lambda^2\right)
e^{-{\bar \alpha}_s 
\ln{\frac{\left(\vec{k}_a+\vec{k}\right)^2}{\vec{k}_a^2}}{\rm Y}}
{\tilde f} \left(\vec{k}_a+\vec{k},\vec{k}_b,{\rm Y}\right),
\end{eqnarray}
if the inital condition
\begin{eqnarray}
{\tilde f} \left(\vec{k}_a,\vec{k}_b,{\rm Y}\right) &=& 
\delta^{(2)}\left(\vec{k}_a-\vec{k}_b\right),
\end{eqnarray}
is chosen for a function related to the gluon Green's function up to a 
factor including the Reggeized gluon propagator, {\it i.e.}
\begin{eqnarray}
f \left(\vec{k}_a,\vec{k}_b,{\rm Y}\right) &\equiv&
e^{\omega_0 \left(\vec{k}_a^2,\lambda^2\right) {\rm Y}}
{\tilde f} \left(\vec{k}_a,\vec{k}_b,{\rm Y}\right). 
\end{eqnarray}
The generalization of this formula to the case of colour singlet exchange 
is straightforward~\cite{NFggf} and the evolution equation can be written as 
\begin{eqnarray}
\frac{\partial}{\partial {\rm Y}} 
{\tilde f} \left(\vec{k}_a,\vec{k}_b,\vec{q},{\rm Y}\right) &\equiv& 
\int d^2 \vec{k} \frac{{\bar \alpha}_s }{2\pi \vec{k}^2} \left(1+\frac{\vec{k}_a^{*2} \left(\vec{k}_a+\vec{k}\right)^2-\vec{q}^{\,2} \vec{k}^2}{\vec{k}_a^2\left(\vec{k}_a^*+\vec{k}\right)^2}\right)\theta\left(\vec{k}^2-\lambda^2\right)
\nonumber\\
&\times& e^{-\frac{{\bar \alpha}_s}{2} 
\ln{\frac{\left(\vec{k}_a+\vec{k}\right)^2
\left(\vec{k}_a^*+\vec{k}\right)^2}{\vec{k}_a^2\vec{k}_a^{*2}}}{\rm Y}}
{\tilde f} \left(\vec{k}_a+\vec{k},\vec{k}_b,\vec{q},{\rm Y}\right).
\end{eqnarray}
The generalization for the NLO case has a very similar structure although 
the kernels and gluon Regge trajectories are more complicated. 

\begin{figure}
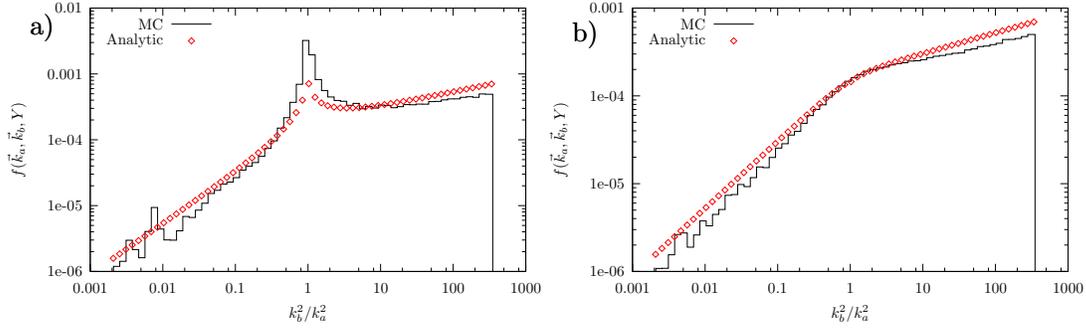

\centering
\epsfig{file=plots.1, width=2.8in}
\epsfig{file=plots.0, width=2.8in}
\caption{Comparison of LO forward analytic versus MC for a fixed angles
a) $\phi = 0$ b) $\phi = \pi/4$. For these figures we take $Y=4$ and 
$\vec{k}_a = (50, 0)$ GeV. 
The source of the discrepancy, in both figures, away
from $k_b^2/k_a^2 = 1$ is still under investigation, as is the ehancement for
$\phi=0$ around $k_b^2/k_a^2 = 1$.
\label{fig:anglef}}
\end{figure}

The numerical results are presented in fig.~
\ref{fig:anglef}. These correspond to the LO forward case with $Y=4$ and $\vec{k}_a
= (50, 0)$ GeV. Work is in progress to implement other cases at LO and NLO as well. 
In these figures the analytic solution is found from the sum of conformal spins up
to 20 and the maximum number of emissions in the Monte Carlo implementation is
30. As can be seen there is still a small discrepancy as the ratio $k_b^2/k_a^2$ 
moves away from 1. This is still under investigation.


One of the questions we would like to answer is what is the effect of 
the collinear resummation in $k_t$--space proposed in Ref.~\cite{Vera:2005jt} 
on the behaviour of the Green's function when the value of the transverse 
momenta in the incoming Reggeons is very different. It is well--known that 
for larger values of the coupling this region presents an instability in 
terms of convergence and it would be important to understand how to solve 
this situation in a fully exclusive way by complementing the BFKL approach 
with all--order contributions dominant in the collinear region. The 
prescription proposed in~\cite{Vera:2005jt} has a simple implementation 
since it suggests that the modification needed in the NLO kernel to introduce 
the collinear improvements is to remove the term
\begin{eqnarray}
-\frac{\bar{\alpha}_s^2}{4}\frac{1}{\pi (\vec{q}-\vec{p})^2}
\ln^2\left({\frac{q^2}{p^2}}\right)
\end{eqnarray}
 in the NLO real emission kernel and replace it with 
\begin{eqnarray}
\frac{1}{\pi (\vec{q}-\vec{p})^2} \left\{\left(\frac{q^2}{p^2}\right)^{-{\rm b}{\bar \alpha}_s 
\frac{\left|p-q\right|}{p-q}}
\sqrt{\frac{2\left({\bar \alpha}_s+ {\rm a} \,{\bar \alpha}_s^2\right)}{\ln^2{\left(\frac{q^2}{p^2}\right)}}} 
J_1 \left(\sqrt{2\left({\bar \alpha}_s+ {\rm a} \,{\bar \alpha}_s^2\right) 
\ln^{2}{\left(\frac{q^2}{p^2}\right)}}\right) \right.\nonumber\\
&& \left.\hspace{-7cm}- {\bar \alpha}_s - {\rm a} \, {\bar \alpha}_s^2
+ {\rm b} \, {\bar \alpha}_s^2 \frac{\left|p-q\right|}{p-q}
\ln{\left(\frac{q^2}{p^2}\right)} \right\}
\end{eqnarray}
where
\begin{eqnarray}
{\rm a} &=& \frac{5}{12}\frac{\beta_0}{N_c} -\frac{13}{36}\frac{n_f}{N_c^3}
-\frac{55}{36}, \, \, \,
{\rm b} ~=~ -\frac{1}{8}\frac{\beta_0}{N_c} -\frac{n_f}{6 N_c^3}
-\frac{11}{12}
\end{eqnarray}
are the collinear coefficients of the kernel in Mellin space.

Recently a theoretical derivation of a jet definition valid at NLO has been 
calculated in the context of the NLO BFKL equation in 
Ref.~\cite{Bartels:2006hg}. One of the targets of our research will be to 
implement this jet definition and study jet multiplicities of fully generated 
events at NLO. Issues related to the conservation of longitudinal components 
in the final state emissions can be studied in this context.

\section{Reggeon Effective Action}
The Reggeon effective action~\cite{Antonov:2004hh} contains a large quantity of 
physics. The problem is how to retrieve the interesting and possibly new physics 
from this action. This is still a new area of work and there is little experience 
about the best path to proceed down. We have decided to develop a computational tool 
to help us understand this action and study its behaviour.

The action is the standard QCD action plus a term induced by the reggeization
of the gluon. This term can be expressed in terms of a kinetic term, describing
the propagation of a Reggeon field, and Reggeon-gluon coupling terms
\begin{equation}
{\cal L}_{ind} = {\cal L}^k_{ind} + {\cal L}^{RG}_{ind}.
\end{equation}

In order to discuss these terms, we must introduce the kinematics implicit
in this action. The action is defined in the quasi--multi--Reggeon kinematics
(QMRK) approximation. In this regime the final states are produced in clusters,
with each cluster having total momentum $Q_i$. Cluster $k$ is composed of any
number of gluons, each containing momentum $p_j^{(k)}$. Therefore we have
\begin{eqnarray}
P_A + P_B &=& Q_1 + Q_2 + \dots + Q_n\, ; \, \, \,  Q_i^2 = M_i^2, \, Q_k =
\sum_j p^{(k)}_j, \\
s &=& 2 P_A P_B = 4 E^2 \gg s_i = 2 Q_i Q_{i+1} \gg |t_i| = |q_i^2|,
\end{eqnarray}
where $P_A$ and $P_B$ are incoming momenta.
We now introduce the light-cone vector
\begin{equation}
n^\mu_\pm = P^\mu_{B/A}/E \, ; \, (n^\pm)^2 = 0;\, \, n^+ \cdot n^- = 2 \, ; \,  
n^\pm \cdot \vec{k}_T = 0.
\end{equation}
We can now use this projection to project the Lorentz derivatives and define the 
inverse projected Lorentz derivative acting on a field
\begin{equation}
\partial_\pm = n^\mu_\pm \partial_\mu \, ; \, \frac{1}{\partial_\pm} A^\mu = \frac{i}
{p_\pm} A^\mu.
\end{equation}

Using these definitions the kinetic term is
\begin{equation}
{\cal L}^k_{ind} = -\partial_+ R^- \partial_- R^+, \label{eqn:kin}
\end{equation}
for Reggeon field $R$. In the QMRK the Reggeon momentum is transverse. This is
given by the condition $\partial_\pm R^\pm = 0$ which is inherently 
included in eqn.~(\ref{eqn:kin}). The induced couplings between Reggeon fields
and gluonic fields is then
\begin{equation}
{\cal L}^{RG}_{ind} = -{\rm Tr} \left\{ \frac{\partial_+}{g} \left[{\cal P} 
\exp \left(-\frac{1}{2} \int_{-\infty}^{x^+} A_+(x') dx'^+ \right) \right]
\partial^2 R_-(x) + (- \leftrightarrow +) \right\}. \label{eqn:action}
\end{equation}
For notational simplicity we have defined
\begin{equation}
A_\mu = -i T^a A^a_\mu, \, R_\pm = -i T^a R_\pm^a,
\end{equation}
where $T^a$ are the $SU(3)$ generators in the fundamental representation.
The trace is over colour indices. This path exponentiated integral can be
written in the form
\begin{equation}
\frac{\partial_+}{g} {\cal P} \exp \left( -\frac{1}{2} \int_{-\infty}^{x^+} 
A_+(x') dx'^+ \right) = A_+ \sum_{i=0}^{\infty} \left( \frac{-g}
{\partial_+} A_+ \right)^i,
\end{equation}
and can then be used to define the interaction vertices of the Reggeon
field with the gluonic fields.

We wish to be able to use the action to calculate some observables. In order to do
this a convenient tool to have is the effective vertices which couple 
$n$ Reggeon fields to $m$ gluonic fields. This requires generating the correct
topologies and then extracting from eqn.~(\ref{eqn:action}) the bare vertices
and applying them to the topology. For this task we are using the
analytic effective action analysis tool {\it Effective}~\cite{Hetherington:2006pa}. 
This tool 
allows us to define the Lagrangian (with the additional definitions of the operators)
and extract the appropriate couplings from this action. We couple this
to a topology generator to produce the tree-level effective vertices
for $n$ Reggeon and $m$ gluons.

\begin{figure}
\centering
\begin{minipage}{2in}
\epsfig{file=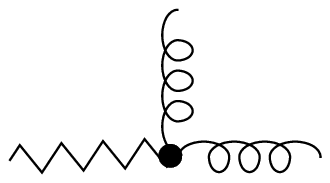}
\put(-50,-20){(T1)}
\put(-100,15){\small $-,c,q$}
\put(-20,15){\small $\mu,a,p_1$}
\put(-38,40){\small $\nu,b,p_2$}
\end{minipage}
\begin{minipage}{2in}
\epsfig{file=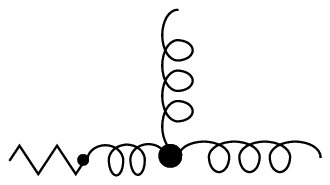} 
\put(-50,-20){(T2)}
\put(-100,15){\small $-,c,q$}
\put(-20,15){\small $\mu,a,p_1$}
\put(-38,40){\small $\nu,b,p_2$}
\end{minipage}
\caption{Topology of a minus Reggeon to two gluons. Zig-zag line is a Reggeon
while swirly lines are gluons.\label{fig:top1}}
\end{figure}

We will give now an example of calculating the effective vertex which couples
a Reggeon with minus light-cone component to two gluons. 
Figure~\ref{fig:top1} shows the topologies of this effective action. We see that 
this has two contributions. The first is the bare vertex which can
be found in the action. We see, in momentum space, this is simply
\begin{equation}
{\cal L}_{ind}^{RGG} = - i q^2 g \left( \frac{A_+(p_1) A_+(p_2) 
R_-(q)}{p_i^+} + \frac{A_-(p_1) A_-(p_2) R^+(q)}{p_i^-} \right),
\end{equation}
where $i$ can be the momentum of either gluon. The colour factor must also be
extracted from these terms. Doing so we find
\begin{equation}
A_+(p_1) A_+(p_2) R_-(q) = (-i)^3 T^a T^b T^c A^+_a(p_1) A^+_b(p_2) R_c^-(q).
\end{equation}
Using the condition $p_1^- + p_2^- = 0$ for the minus Reggeon and
\begin{equation}
A^\mu = \frac{1}{2} (n^-)^\mu A^+ + \frac{1}{2} (n^+)^\mu A^- + A^T,
\end{equation}
we can sum the two contributions (choice of $i$ in momentum) to find
\begin{equation}
\left<0 \left| {\cal L}_{ind}^{RG} \right| A_a^\mu(p_1) A_b^\nu(p_2) R_c^-(q)
\right> = i g q^2 f_{abc} (n^-)^\mu (n^-)^\nu \frac{1}{p_1^-}. \label{eqn:p1}
\end{equation}
For the other topology we have a Reggeon-gluon transition which can be found
from the ${\cal O}(g^0)$ term in the induced action, and a standard QCD 
three-gluon vertex. The transition term simply provides a coupling
$-q^2 \frac{\delta^{aa'}}{2} (n^-)^\lambda$. The three-gluon coupling vertex is
\begin{equation}
g f_{abc} \left[ (q - p_1)_\nu g_{\lambda \mu} +
(p_1-p_2)_\lambda g_{\mu \nu} + (p_2 - q)_\mu g_{\nu \lambda} \right].
\end{equation}
Using these we find the contribution from the (T2) topology
\begin{equation}
g f_{abc} \left[ (p_1 - p_2)^- g^{\mu \nu} + 
(-2 p_1 - p_2)^\nu (n^-)^\mu + (2 p_2 + p_1)^\mu (n^-)^\nu \right],
\end{equation}
which can be combined with eqn.~(\ref{eqn:p1}) to give the full effective 
vertex
\begin{equation}
V^{\mu \nu -}_{a b c} = g f_{abc} \left[ (p_1 - p_2)^- g^{\mu \nu}
+ (-2 p_1 - p_2)^\nu (n^-)^\mu + (2p_2 + p_1)^\mu (n^-)^\nu - 
\frac{q^2}{p_1^-} (n^-)^\mu (n^-)^\nu
\right].
\end{equation}

The software package being implemented will calculate the effective vertices in 
the same manner as what is presented here. We plan to use our code to verify the 
other vertices found in~\cite{Antonov:2004hh}, 
rather than calculating them by hand. We also plan to calculate new vertices
which have not been presented in~\cite{Antonov:2004hh}. Using an automated system
has the additional advantage that it may allow us to integrate with other codes, 
e.g. matrix element generators.

\section{Outlook}
We have presented the preliminary work on two computational tools. The first
is a Monte Carlo to compute the solution to the BFKL equation for several 
different kernels. We also want this code to be able to produce fully 
exclusive events at NLO, but this requires still a lot of work. We have shown
the results of our current implementation for the LO forward kernel and 
discussed the other kernels which are of interest to us.

The second part of these proceedings introduces another computational tool
designed to begin the analysis of the effective action proposed 
in~\cite{Antonov:2004hh}. 
This action is quite complicated, but may include important and interesting physics. 
Because of the complexity to analyze the action, we feel a computational tool
is ideally suited for this purpose. Thus we introduce the code which we are
currently writing to analyze this action. We also present an example of the
type of calculation which this code will perform.

\noindent
{\bf Acknowledgments:} 
Discussions with J.~Bartels and L.~Lipatov are acknowledged.  PS would 
also like to thank the conference organizers for their invitation and 
financial support.


\begin{thebibliography}{99}

\bibitem{FKL}  L.\thinspace N.~Lipatov, Sov.\ J.\ Nucl.\ Phys.\ {\bf 23}, 338 (1976); V.\thinspace S.~Fadin, E.\thinspace A.~Kuraev and L.\thinspace N.~Lipatov, Phys.\ Lett.\ B {\bf 60}, 50 (1975), Sov.\ Phys.\ JETP {\bf 44}, 443 (1976), Sov.\ Phys.\ JETP {\bf 45}, 199 (1977); I.\thinspace I.~Balitsky and L.\thinspace
N.~Lipatov, Sov.\ J.\ Nucl.\ Phys.\ {\bf 28}, 822 (1978), JETP\ Lett.\ {\bf 30}, 355 (1979).

\bibitem{FLCC}
  V.S.~Fadin, L.N.~Lipatov, Phys.\ Lett.\ B {\bf 429}, 127 (1998);
  G.~Camici, M.~Ciafaloni, Phys.\ Lett.\ B {\bf 430}, 349 (1998).               
\bibitem{MC-GGF}
  J.~R.~Andersen, A.~Sabio Vera,  Phys.\ Lett.\ B {\bf 567}, 116 (2003);
                                  Nucl.\ Phys.\ B {\bf 679}, 345 (2004);
                                  Nucl.\ Phys.\ B {\bf 699}, 90 (2004);
                                  JHEP {\bf 0501}, 045 (2005).
                                  

\bibitem{JRA}
  J.~R.~Andersen,  Phys.\ Lett.\ B {\bf 639}, 290 (2006).

\bibitem{Lipatov:1995pn}
  L.~N.~Lipatov,  Nucl.\ Phys.\ B {\bf 452}, 369 (1995).

\bibitem{Schmidt:1996fg}
  C.~R.~Schmidt,  Phys.\ Rev.\ Lett.\  {\bf 78}, 4531 (1997).

\bibitem{NFggf}
  J.~R.~Andersen and A.~Sabio~Vera, JHEP {\bf 0501}, 045 (2005).

\bibitem{Vera:2005jt}
  A.~Sabio~Vera,  Nucl.\ Phys.\ B {\bf 722}, 65 (2005).


\bibitem{Vera:2006un}
  A.~Sabio~Vera,  Nucl.\ Phys.\ B {\bf 746}, 1 (2006).

\bibitem{Bartels:2006hg}
  J.~Bartels, A.~Sabio Vera and F.~Schwennsen, hep-ph/0608154.

\bibitem{Antonov:2004hh}
  E.~N.~Antonov, L.~N.~Lipatov, E.~A.~Kuraev and I.~O.~Cherednikov,
  Nucl.\ Phys.\ B {\bf 721}, 111 (2005)
  [arXiv:hep-ph/0411185].

\bibitem{Hetherington:2006pa}
  J.~P.~J.~Hetherington and P.~Stephens,
  arXiv:hep-ph/0605149.


\end{thebibliography}
\end{document}